\documentstyle[epsf,12pt]{article}

\begin{document}
\baselineskip 4ex
\begin{center}
{\Large\bf Low-Energy \mbox{\boldmath $\Lambda\!-\!p$}
{\bf Scattering Parameters from}} \\[1ex] 
{\Large\bf the \mbox{\boldmath $pp\rightarrow pK^+\Lambda$}~{\bf Reaction }}
\\[7ex]
\end{center}

\noindent
 J.T.~Balewski$^{1, 2}$,
 A.~Budzanowski$^1$,
 C.~Goodman$^3$,
 D.~Grzonka$^2$,
 M.~Hofmann$^2$,
 L.~Jarczyk$^4$,
 A.~Khoukaz$^5$,
 K.~Kilian$^2$,
 T.~Lister$^5$,
 P.~Moskal$^{2, 4}$,
 W.~Oelert$^2$,
 I.A.~Pellmann$^{2, 5}$,
 C.~Quentmeier$^5$,
 R.~Santo$^5$,
 G.~Schepers$^5$,
 T.~Sefzick$^2$,
 S.~Sewerin$^2$,
 J.~Smyrski$^4$,
 A.~Strza\l kowski$^4$,
 C.~Thomas$^5$,
 C.~Wilkin$^6$,
 M.~Wolke$^2$,
 P.~W\"ustner$^7$,
 D.~Wyrwa$^{2, 4}$\\[2ex]

\noindent
\textit{$^1$ Institute of Nuclear Physics, Cracow, Poland }\\
\textit{$^2$ IKP, Forschungszentrum J\"ulich, Germany }\\
\textit{$^3$ IUCF, Bloomington, Indiana, USA } \\
\textit{$^4$ Institute of Physics, Jagellonian University, Cracow, Poland }\\
\textit{$^5$ IKP, Westf\"alische Wilhelms--Universit\"at, M\"unster, Germany
}\\
\textit{$^6$ University College London, London WC1E 6BT, United Kingdom }\\
\textit{$^7$ ZEL, Forschungszentrum J\"ulich, Germany } 

\begin{center}
\end{center}

\begin{abstract}
Constraints on the spin-averaged $\Lambda p$ scattering length and effective
range have been obtained from measurements of the 
$pp\rightarrow pK^+\Lambda$ reaction close to the production threshold
by comparing model phase-space Dalitz plot occupations with experimental ones.
The data fix well the position of the virtual bound state in the
$\Lambda p$ system. Combining this with information from elastic 
$\Lambda p$ scattering measurements at slightly higher energies,
together with the fact that the hyperdeuteron is not bound, leads to
a new determination of the low energy $\Lambda p$ scattering parameters.\\[3ex]
\end{abstract}


\newpage
\baselineskip 4ex
\section{Introduction }
The existence of light hypernuclei, such as $_{\Lambda}^{3}$He, shows the
low energy $\Lambda$-$p$ interaction to be strongly 
attractive, though not sufficient to bind the two-baryon hyperdeuteron.
The $\Lambda$-$p$ interaction is of especial interest since it is influenced 
by the  strange quark content of the $\Lambda$-hyperon. However, in contrast 
to the nucleon-nucleon case, due to the short lifetime of the $\Lambda$, 
direct measurements of low-energy $\Lambda$-$p$ scattering are sparse and the 
resulting parameters rather poorly known.

\begin{figure}[htb]
\epsfysize=10.cm \centering \leavevmode \epsfverbosetrue \epsfclipon
\epsffile[38 501 275 657 ]{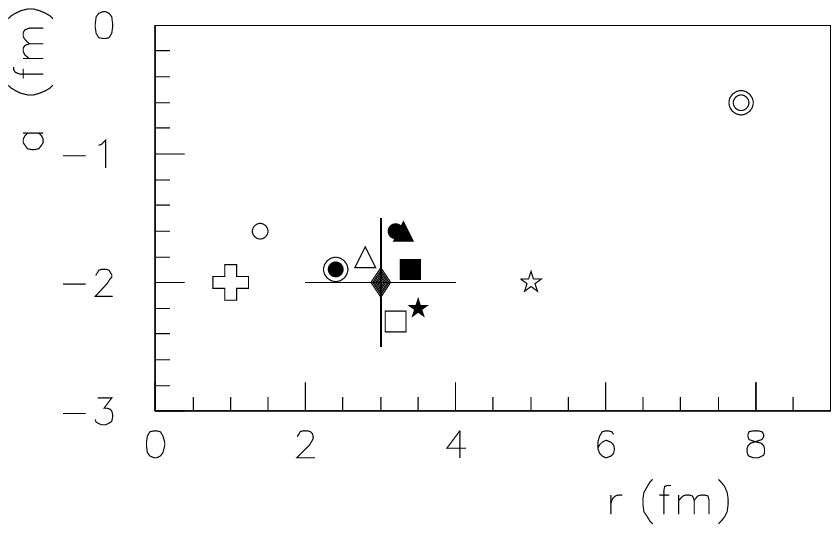}
\caption{$\Lambda$-$p$~scattering parameters for the singlet and triplet states
 marked with open and closed symbols, respectively. Values obtained from 
 experimental $\Lambda$-$p$ elastic scattering data of Refs.~[1] (stars) and [2] 
 (triangles) are shown as well as the triplet values obtained from a $K^-$ 
 capture experiment [4] (diamond). It is only in this latter case that an 
 attempt was made to quote errors which are, however, strongly correlated. 
 Points deduced from the phenomenological potential models of Refs.~[5] with 
 solution A (circles) and solution B (circles with additional outer circle) 
 and [6] (squares) are also shown. }
\label{ar_other}
\end{figure}

Bubble chamber measurements \cite{Alex68,Sech68,Alex69}, based on samples of a 
few hundred secondary events, have allowed determinations of the elastic
cross section down to $\Lambda$ laboratory momenta $\approx 130$~MeV/c. 
In the low energy region, where only $S$-waves are important, the 
spin-averaged total cross section is of the form
\begin{equation}
\sigma_{\Lambda p\to \Lambda p} = \frac{\pi}{q^2+\left(-1/a_s+
\mbox{${\textstyle \frac{1}{2}}$}r_s q^2\right)^2}
+ \frac{3\pi}{q^2+\left(-1/a_t+
\mbox{${\textstyle \frac{1}{2}}$}r_t q^2\right)^2}\:\cdot
\end{equation}
Here $q$ is the $\Lambda p$ centre-of-mass momentum and $a_{s(t)}$ and
$r_{s(t)}$ are, respectively, the $S$-wave scattering lengths and effective 
ranges in the $\Lambda p$ spin-singlet and triplet states. Separate values 
of these parameters have been claimed for the two spin states 
\cite{Alex68,Sech68} and these are shown in Fig.~\ref{ar_other}. However, the error bars 
are large, strongly and systematically
correlated and hard to quantify, since such data should 
really only support the determination of an average scattering length 
$\bar{a}$ and effective range $\bar{r}$ \cite{Alex69}. Already for laboratory
momenta $\approx 300$~MeV/c, the differential cross section is significantly 
non-isotropic, indicating the presence of $P$ or higher waves 
\cite{Alex68,Sech68}, and so it is not surprising that the $S$-wave parameters 
deduced from such experiments depend upon the upper momentum cut assumed.

Values of the scattering length and effective range have also been deduced
through the study of the $\Lambda p$ final state interaction (FSI) in the
$K^-d\to \pi^-p\Lambda$ reaction with stopped $K$-mesons \cite{Tan}. Here it
is the \underline{shape} of the $\Lambda p$ effective mass spectrum near the 
kinematic limit which is sensitive to the parameters. In impulse approximation
the amplitude for this process is proportional to that for 
$K^-n\to \pi^-\Lambda$ and, if the Fermi motion in the target deuteron is 
neglected, the reaction is purely $s$-wave with  no spin-flip.
The final $\Lambda p$ system is therefore in the same spin-triplet
state as the $np$ pair in the deuteron and the values $a_t$ and $r_t$ so 
determined, which are also shown in Fig.~\ref{ar_other}, are consistent with 
those obtained from the scattering experiments \cite{Alex68,Sech68}.

Phenomenological investigations of the $\Lambda p$ interaction by
the J\"ulich~\cite{Holz89} and Nijmegen~\cite{Nij} groups yield low energy
scattering parameters in agreement with the results of Fig.~\ref{ar_other}, 
though it is
impossible to quantify the systematic uncertainties inherent in such models.
It should, however, be noted that their spin-singlet potential is more 
attractive than the triplet, which is necessary to ensure the correct spin 
assignments of the ground states of $_{\Lambda}^{3}$H and $_{\Lambda}^{4}$He 
\cite{Dalitz}.

The final state interaction in the $pp\to K^+p\Lambda$ reaction at low
$\Lambda p$ effective masses is also sensitive to the $\Lambda p$ scattering 
parameters \cite{Saturne}. The recent data on this reaction very close
to threshold~\cite{Bal97}, taken at the COSY-11 magnetic spectrometer 
\cite{Bra96} installed at the cooler synchrotron COSY-J\"ulich~\cite{COSY}, 
allow us to extract information
complementary to that obtained from elastic 
scattering because it is possible to reach lower centre-of-mass momenta.

In the present paper we aim to constrain the $\Lambda p$ scattering
parameters from the \underline{shapes} of the double-differential  
$pp\to K^+p\Lambda$ cross section. In section~2 a simplified model is
outlined to describe the principal interactions between the three outgoing 
particles. Experimental details, including event selection, are
discussed in section~3.
The FSI model was already used in ref.~\cite{Bal97} to determine the
precise beam energy by fitting the integrated total cross sections as
a function of beam energy with assumed values of the $\Lambda p$ input
parameters. To avoid biassing the analysis, in the present work we determine 
the $\Lambda p$ parameters using only the structure of the Dalitz plot at 
each energy and not its normalisation. For this purpose we apply the maximum 
likelihood method to obtain a map of the confidence levels for the
$\Lambda p$ scattering  parameters. This formalism and the definition of the
event weights are shown in section~4. Even taking the $\Lambda p$ spin-triplet 
and singlet to be identical, the resulting values of the average
scattering length and effective range presented in section~5 are strongly and
systematically
correlated in the fit, such that it is hard to quote error bars. However, the 
position of the nearby pole in the $\Lambda p$ scattering amplitude, 
corresponding to a virtual bound state of the system, is much more stable,
being unbound by $(7.7^{+6.0}_{-3.0})$~MeV. Our conclusions are presented in
section 6.

\section{Model for the FSI in the \mbox{\boldmath $pp\rightarrow pK^+\Lambda$}
reaction}

As discussed in ref.~\cite{Bal97}, if the basic production mechanism is of
short range then the energy dependence of the $pp\rightarrow pK^+\Lambda$ 
cross section close to threshold is dominated by the available three-body 
phase-space $d\rho (3)$, modified by final-state interactions. In principle 
one should consider FSI's in all the three two-body subsystems, 
\mbox{$\Lambda$-$p$}, \mbox{$\Lambda$-$K$}, and \mbox{$p$-$K$}. 
Since the strong 
interaction in the first case appears to be more than an order of
magnitude larger than for the other two \cite {Hoff,Deloff}, we concentrate on
the dominant factor $f_{_{FSI}}(q)$ in the $\Lambda$-$p$ system.
In addition, however, the Coulomb repulsion in the proton-kaon
subsystem $f_c(q_{pK})$, where $q_{pK}$ denotes the c.m.\ momentum in $pK$ 
subsystem, is also important at the low energies pertaining 
in our experiment. Keeping only these two interactions our {\it ansatz} for 
the production cross section is
\begin{eqnarray}
d\sigma
~~~\sim~~~  f_c(q_{p_K})\: f_{FSI}(q) \: d\rho (3)\:.
\label{eq_Iffr}
\end{eqnarray}

The $S$-wave assumption implicit here is justified for our 
experiment since, for excess energies $\varepsilon <7$~MeV, the maximum
momentum in any two-body subsystem is below 100~MeV/c. It should be noted
that the bubble chamber data \cite{Alex68,Sech68} cover a higher range of 
$\Lambda$-$p$ momenta.

Choosing as independent variables $S_{pK}$ and $S_{\Lambda K}$,
the squares of the effective masses in the $p$-$K$ and 
$\Lambda$-$K$ subsystems, integration of Eq.~(\ref{eq_Iffr}) over the angular 
variables leads to a number of events per pixel in the Dalitz plot 
distribution of the form
\begin{eqnarray}
\displaystyle
\frac{d^2\sigma}{dS_{pK}~ dS_{\Lambda K}}    ~~~\sim~
  \int_{\mbox{\tiny angles}}\  
\!\! f_c(q_{p_K}) \: f_{_{FSI}}(q)\: d\rho (3) \;.
\label{eq_dffr}
\end{eqnarray}

\begin{figure}[htb]
\epsfysize=12.cm \centering \leavevmode \epsfverbosetrue \epsfclipon
\epsffile[25 350 278 600 ]{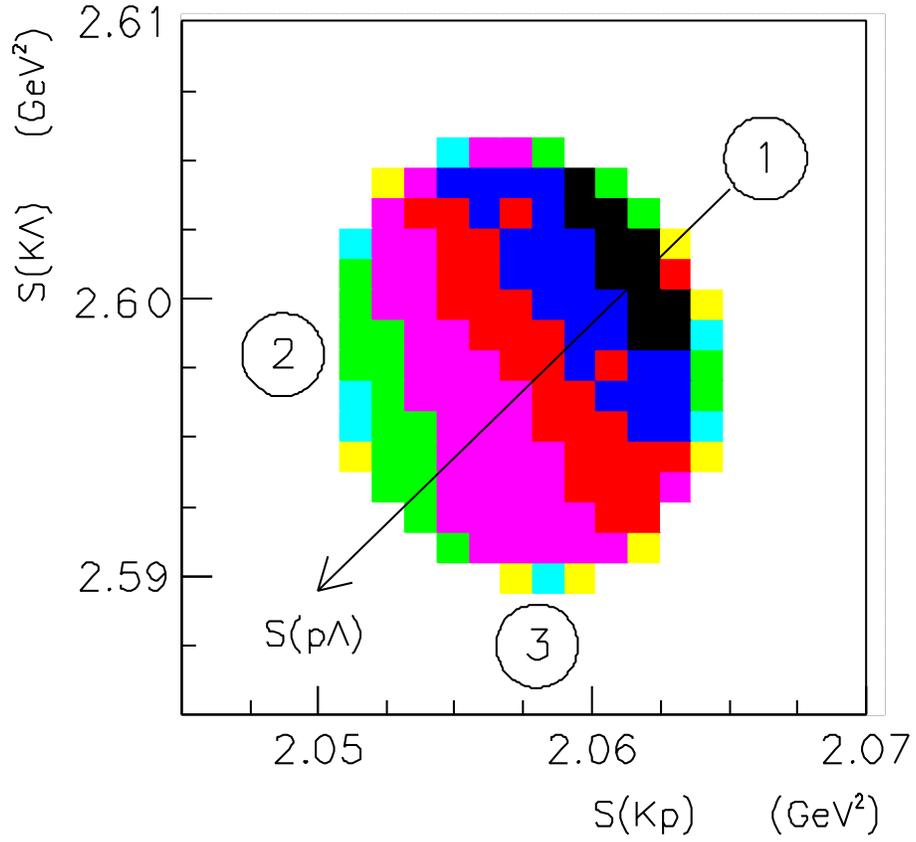}

\caption{Monte-Carlo Dalitz plot for the $pp\to p K^{+}\Lambda$ reaction
 at an excess energy $\varepsilon=4.7~MeV$.
 The numbers 1, 2, and 3 mark regions of diminishing
 relative energy in the $\Lambda p$, $K^+p$, and $K^+\Lambda$ two-body
 systems, respectively.}
\label{dal_mod_q5_new}
\end{figure}

Structure in the Dalitz plot must be associated with the functions $f_{FSI}$ 
and $f_c$. Now the Coulomb distortion factor defined in Ref.~\cite{Bal97}
contains no free parameters \cite{Hanh}. In contrast the $f_{FSI}$ factor
depends on the scattering lengths and effective ranges in the triplet
and singlet states. Since the spin dependence is expected to be 
small~\cite{Holz89,Nij}, as shown by some of the extracted numbers 
in Fig.~\ref{ar_other}, and our experiment is not sensitive to singlet/triplet differences,
we used mean values of the scattering length \mbox{$\bar{a}$} and
effective range \mbox{$\bar{r}$} in the parametrization of $f_{FSI}$.
In analogy with Eq.~(1), we take the popular Watson form for the final state
interaction \cite{GW}
\begin{eqnarray}
\displaystyle
 f_{_{FSI}}(q,\bar{a},\bar{r})  = \frac{1}{\bar{a}^2q^2+
\left(-1+\mbox{${\textstyle \frac{1}{2}}$}\bar{r}\bar{a}q^2\right)^{\!2}}
\:\cdot
\label{eq_fsi}
\end{eqnarray}
A typical  Dalitz plot distribution calculated from Eqs.~(\ref{eq_dffr}) and
(\ref{eq_fsi}) with $\bar{a} =-1.6$~fm and $\bar{r}=2.3$~fm is
shown in Fig.~\ref{dal_mod_q5_new}.

\section{Experiment}

The measurement of the $pp\rightarrow pK^+\Lambda$ reaction was performed 
at the COSY-J\"ulich synchrotron, using the internal target facility 
COSY-11 \cite{Bra96}. Outgoing protons and positively charged kaons were 
identified by means of particle momentum reconstruction in the magnetic 
field combined with time-of-flight measurement. The four-momentum, and
hence the missing mass (MM), corresponding to the unobserved $\Lambda$-hyperon 
was calculated from energy-momentum conservation. Details of the experimental 
technique are given elsewhere~\cite{Bal96}.

\begin{figure}[htb]
\epsfysize=12.cm \centering \leavevmode \epsfverbosetrue \epsfclipon
\epsffile[27 428 299 661 ]{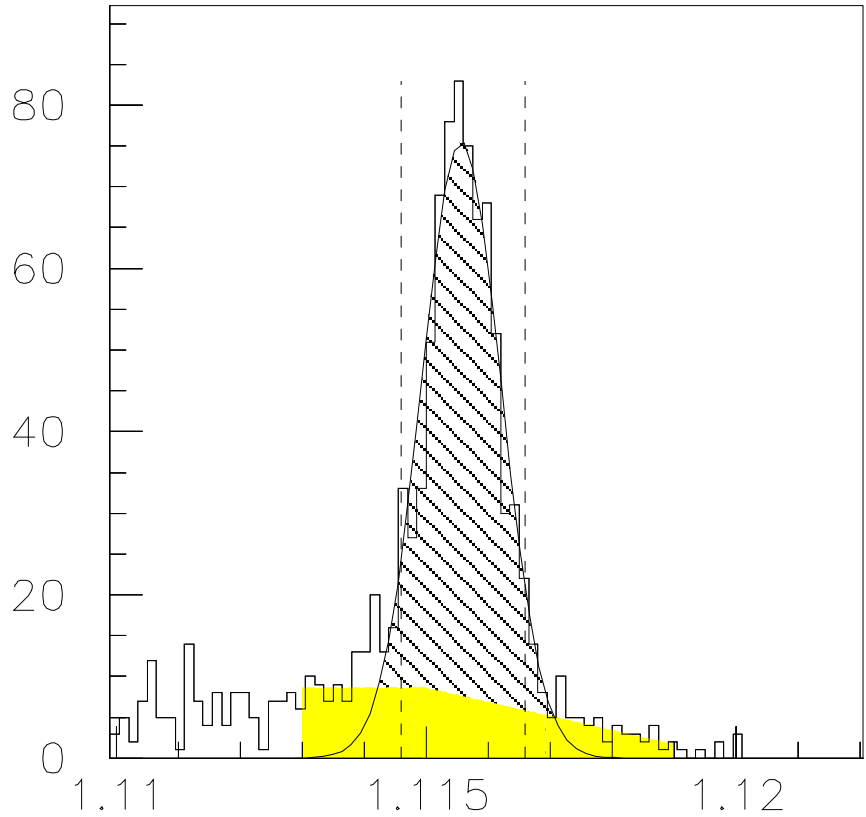}

\caption{
Experimental missing mass distribution for the pK$^+$ subsystem from the 
$pp\to p K^+ X$ reaction. The fitted peak corresponds to the  
$\Lambda$-particle from the $pp\rightarrow pK^+\Lambda$ reaction at an
excess energy of $\varepsilon=4.7~$MeV. The grey area is an estimate of  
the background. Only events within the $\pm2\sigma$ band (dashed lines)
were accepted in the final analysis.}
\label{mm_q5}
\end{figure}

An example of the missing mass spectrum is shown in Fig.~\ref{mm_q5}. The function 
$G(M\!M)$ used to fit the peak corresponding to good
$pp\rightarrow pK^+\Lambda$ events, is combined with the smooth background 
$B(M\!M)$. Only events which deviate by less than $\pm2\sigma$ from the 
central value were accepted and a weight $w={G(M\!M)}/{[G(M\!M)+B(M\!M)]}$
was assigned to each of them to describe the probability that the particular 
event resulted from a $pK^+\Lambda$ final state rather than being a background 
signal. Our data were well described by a Gaussian form for $G(M\!M)$ with
$\sigma=0.5$~MeV. The closer $MM$ is to the known $\Lambda$ mass
$m_\Lambda$, the larger the weight $w$
for the event. Thus $w$ can be interpreted as a penalty factor which is 
imposed on events where the observables are significantly modified by the 
application of the kinematic fit procedure.

\begin{figure}[htb]
\epsfysize=8.cm \centering \leavevmode \epsfverbosetrue \epsfclipon
\epsffile[15 330 520 600 ]{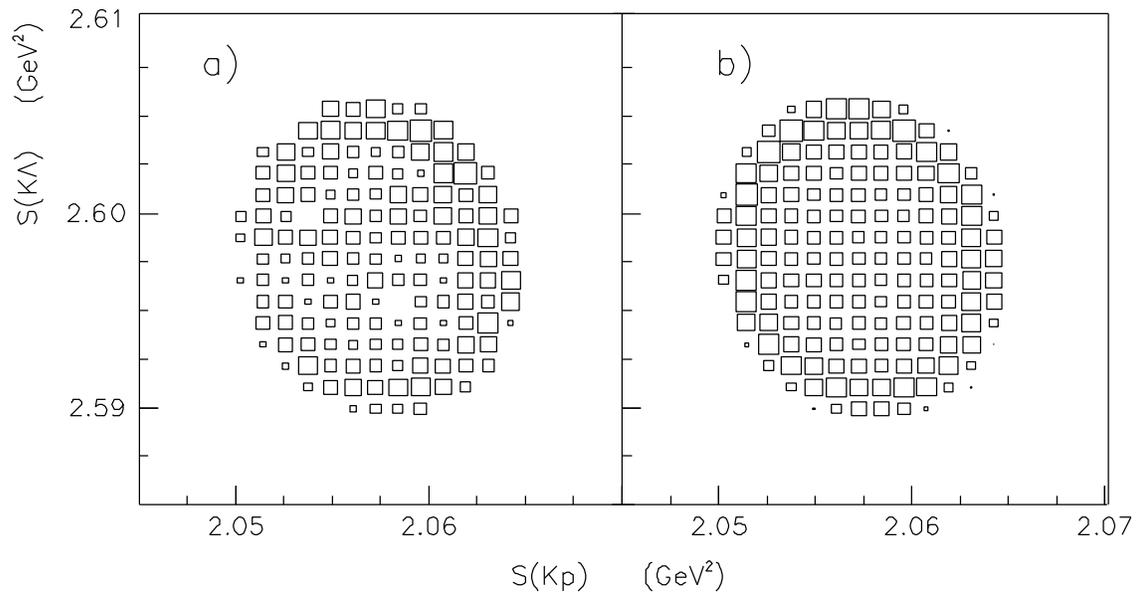}

\caption{a) Experimental Dalitz plot at $\varepsilon=4.7$~MeV containing
776 events;\ 
b) The acceptance of the COSY-11 apparatus shows a typical 
{\it Colosseum}-like structure.}

\label{dal_exp_q5}
\end{figure}

The experimental Dalitz plot for $\varepsilon=4.7$~MeV, shown in
Fig.~\ref{dal_exp_q5}a, 
demonstrates that the whole kinematically allowed region is occupied by data,
so that there are no forbidden zones in our acceptance. This experimental 
acceptance, calculated {\it via} a Monte-Carlo simulation with a pure 
phase-space generator, is however non-uniform with a  {\it Colosseum}-like 
pattern, as illustrated in Fig.~\ref{dal_exp_q5}b. Comparison of the experimental 
Dalitz plot (Fig.~\ref{dal_exp_q5}a) with the model calculation 
(Fig.~\ref{dal_mod_q5_new}), folded with the detector acceptance 
(Fig.~\ref{dal_exp_q5}b), allows a determination of the model parameters
$\bar{a}$ and $\bar{r}$ at each excess energy $\varepsilon$. Although
they influence the statistical confidence, the relative counting rates at 
different $\varepsilon$ are not used, which is important since the
variation of cross section with energy has already been employed in
fixing the absolute beam energy \cite{Bal97}.

\section{Fitting procedure}

The maximum likelihood method was applied to determine best values of the
$\bar{a}$ and $\bar{r}$ parameters, though it must be stressed that these 
will be strongly correlated due to the form of the $f_{FSI}$ factor
of Eq.~(4). A set of model Dalitz plots  
$D_{\bar{a},\bar{r}}(S_{pK},S_{\Lambda K})$,
as defined in Eq.~(\ref{eq_dffr}) and (\ref{eq_fsi}), was generated at fixed 
excess energy $\varepsilon$ over a grid in $(\bar{a},\,\bar{r})$ 
with $\bar{a}\in[-6,~6]$~fm and $\bar{r}\in[0,~10]$~fm and a step-size of 
$0.2$~fm in each variable.

Defining for brevity $x\equiv S_{pK}$ and $y\equiv S_{\Lambda K}$,
the probability $\cal{P}$ that an event will be detected with some $(x,y)$ 
value is
\begin{eqnarray}
\displaystyle {
{\cal{P}}(x,y)=  \left.\vphantom{\int}
{D_{\mbox{$\bar{a},\bar{r}$}}(x,y)\, A(x,y)}\:\right/
{\int \limits_{\mbox{\tiny Dalitz plot}} 
D_{\bar{a},\bar{r}}(x,y)\, A(x,y)\:dx\:dy}}\:,
\label{eq_p1}
\end{eqnarray}
where $A(x,y)$ is the acceptance function.

Applying the experimental event weight $w_i$, the likelihood of a single event
is
\begin{eqnarray*}
\displaystyle
{\textit{l}}_{\,i}= \left[\mbox{$\cal{P}$}(x_i,y_i) \right] ^{w_i}
~\equiv~\mbox{$\cal{P}$} _i^{w_i}
\end{eqnarray*}
and the corresponding global likelihood function 
\begin{eqnarray}
\displaystyle
{\cal{L}}({\mbox{$\bar{a},\bar{r}$}})=\prod \limits_{i=1}^{N}
\mbox{$\cal{P}$}_i^{w_i}\:.
\label{eq_lht}
\end{eqnarray}

Since $w_i$ is the probability that the $i$-th event is a $pK^+\Lambda$
and not a background reaction, if the whole experiment were repeated 
$M$ times with perfect background subtraction, then each $i$-th event would 
appear $M\,w_i$ times. The likelihood corresponding to each
$(x_i,y_i)$ point is given by $\textit{l}_{\,i}^{\,*}= {\cal{P}}_i^{M\,w_i} $ 
and the global likelihood function is:
${\cal{L}}^*=\prod \limits_{i=1}^{N} (\mbox{$\cal{P}$}_i)^{M \,w_i}=
({\cal{L}})^M$.

Both likelihood functions ${\cal{L}}(\bar{a},\bar{r})$ and 
${\cal{L}}^{*}(\bar{a},\bar{r})$ have their extrema at the same
$(\bar{a}_0,\,\bar{r}_0)$ points and thus it is sufficient to perform the
experiment once, searching for the maximum of the ${\cal{L}}(\bar{a},\bar{r})$
function. 

In order to amalgamate the likelihood functions from measurements
at different values of $\varepsilon$, the $M$ was chosen in such a way that
$M\, \sum w_i$ was equal to the number of measured events $N$.

As shown by Eadie~\cite{Eadie}, the quantity
$ -2\,ln\left[{\cal{L}}(\mbox{$\bar{a},\bar{r}$})
/{\cal{L}}(\mbox{$\bar{a}_0,\bar{r}_0$})\right]$
has an asymptotic $\chi^2$ distribution corresponding to the two degrees of 
freedom, {\it i.e.}\ two parameters $(\bar{a},\,\bar{r})$.
The equation
\begin{eqnarray}
\displaystyle
ln~{\cal{L}}(\bar{a},\bar{r}) = ln~{\cal{L}}(\bar{a}_0,\bar{r}_0)- 
\mbox{${\textstyle \frac{1}{2}}$}\chi^2_\beta(2)
\label{eq_cl}
\end{eqnarray}
defines two-dimensional contours in the $(\bar{a},\,\bar{r})$ plane for the 
desired confidence level CL $=1-\beta$.

\section{Results}

\begin{figure}[htb]
\epsfysize=14.cm \centering \leavevmode \epsfverbosetrue \epsfclipon
\epsffile[23 413 276 649 ]{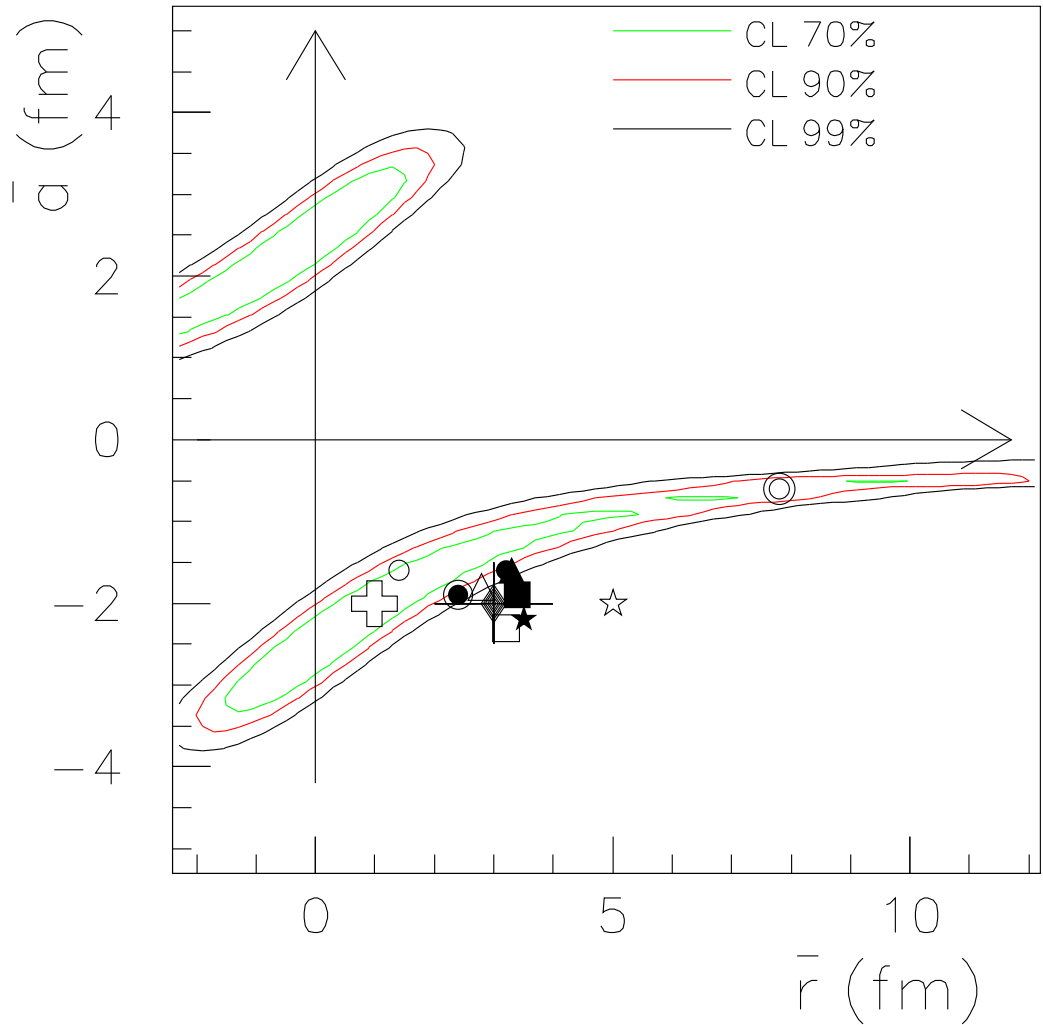}
\caption{The logarithm of the global likelihood function
${\cal{L}}(\bar{a},\bar{r})$ in the $(\bar{a},\,\bar{r})$ plane obtained
using all COSY-11 data measured at $\varepsilon$ between $2.7$ and $6.7$~MeV.
Experimental and theoretical values of these parameters are displayed
using the same symbols as in Fig.~\ref{ar_other}. Combining our data with 
the elastic scattering data of Ref.~[1,2] leads to the open cross at 
$\bar{a}=-2.0$~fm and $\bar{r}=1.0$~fm.}
\label{cl_all_q}
\end{figure}

Contours in the $(\bar{a},\,\bar{r})$ plane of the global likelihood 
function calculated from a sample of about 2400 events measured at six
excess energies from 2.7 to 6.7~MeV are shown in Fig.~\ref{cl_all_q}. Those 
solutions with a positive value of $\bar{a}$ are to be excluded since they 
would imply the existence of a bound $\Lambda p$ system and the hyperdeuteron 
has never been found. The branch with negative $\bar{a}$ yields
a long narrow ridge with a very strong correlation between the
scattering length and effective range, such that it is only a
combination of the two parameters which is well determined by the experiment.
The averaged values from the literature \cite{Alex68}--\cite{Nij} lie
close to the ridge but generally slightly outside the 99\%-confidence contour.

\begin{figure}[htb]
\epsfysize=7.5cm \centering \leavevmode \epsfverbosetrue \epsfclipon
\epsffile[24 347 530 608 ]{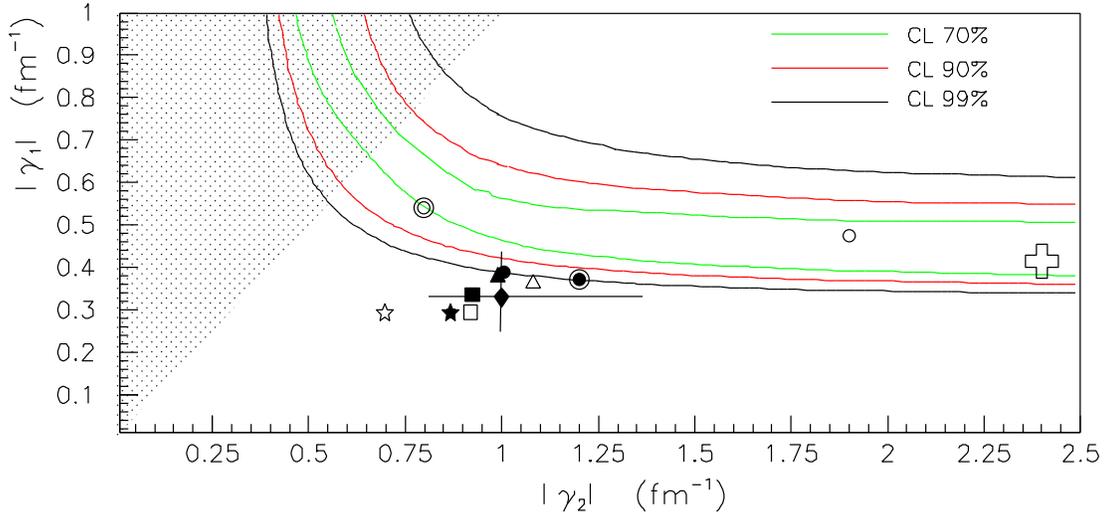}

\caption{The global likelihood function ${\cal{L}}(\gamma_1,\gamma_2)$
transformed into the new variables $(\gamma_1,\gamma_2)$. The literature 
values [1-6] are shown using the same conventions as in Fig.~\ref{ar_other}. 
Since the fit function is symmetric in $(\gamma_1,\,\gamma_2)$ values, only
the curves with $\gamma_2 > \gamma_1$ are significant, and the data
are insensitive to the sign of the $\gamma_n$. The cross at
$(0.41,\, 2.4)$ results from combining our data with that of low
energy elastic $\Lambda p$ scattering [1,2].}
\label{cl_gg}
\end{figure}

The parameter which is in fact well determined by this experiment is
the energy of the nearby pole in the $\Lambda p$ scattering amplitude. 
To see this, rewrite the final-state-interaction factor of Eq.~(4) in the form
\begin{eqnarray}
\displaystyle
f_{FSI}(q;\bar{a},\bar{r}) =\frac{\gamma_1^2\:\gamma_2^2}
{(q^2 + \gamma_1^2)(q^2 + \gamma_2^2)}\:\cdot
\label{eq_xx1}
\end{eqnarray}
\\
The $\Lambda p$ scattering amplitude has poles at $q=i\gamma_n$, where

\begin{eqnarray}
\gamma_1= \frac{1}{\bar{r}}\left[1 -
\sqrt{1-\frac{2\bar{r}}{\bar{a}}}~\right]\ \ \mbox{\rm and}\ \ 
\gamma_2=\frac{1}{\bar{r}}\left[1 + \sqrt{1-\frac{2\bar{r}}{\bar{a}}}~\right]
\:\cdot
\label{eq_xx2}
\end{eqnarray}
\\
It is straightforward from Eq.~(\ref{eq_xx2}) to transform the
likelihood function into the new variables and the contours in the resulting
${\cal{L}}(\gamma_1,\gamma_2)$ are shown in Fig.~\ref{cl_gg} together
with the literature values \cite{Alex68}--\cite{Nij}.
Note that the functional form of Eq.~(\ref{eq_xx1}) clearly shows that
the data are sensitive only to the magnitudes of the $\gamma_i$ and
that the results must be symmetric under the interchange $\gamma_1 
\leftrightarrow \gamma_2$. We therefore establish the convention that
$|\gamma_2| > |\gamma_1|$. If $\gamma_1$ is positive then that would
correspond to a bound state of the $\Lambda p$ system, whereas if it
is negative then it is an antibound or virtual state of the kind with
which one is familiar from the low energy proton-proton singlet $S$-wave.
The two branches seen in the maximum likelihood contours of Fig.~\ref{cl_all_q}
correspond to the mere reversal of the sign of $\gamma_1$. Since it is
highly unlikely that there would be another singularity of the
$\Lambda$-$p$ amplitude very close to zero energy, we can assume that
$|\gamma_2| \gg |\gamma_1|$, in which case we deduce from our fit
that $\gamma_1=(-0.45^{+0.15}_{-0.1})$~fm$^{-1}$. It has been argued \cite{FW}
that the single pole limit of letting $\gamma_2\to\infty$ in fact provides a 
better representation of $f_{FSI}$ than the scattering length -- effective
range form of Eq.~(\ref{eq_fsi}).
Our value of $\gamma_1$ corresponds to the average $\Lambda$-$p$ system being 
unbound by an amount $(7.7^{+6.0}_{-3.0})$~MeV. 

\begin{figure}[htb]
\epsfysize=12.cm \centering \leavevmode \epsfverbosetrue \epsfclipon
\epsffile[24 400 286 658 ]{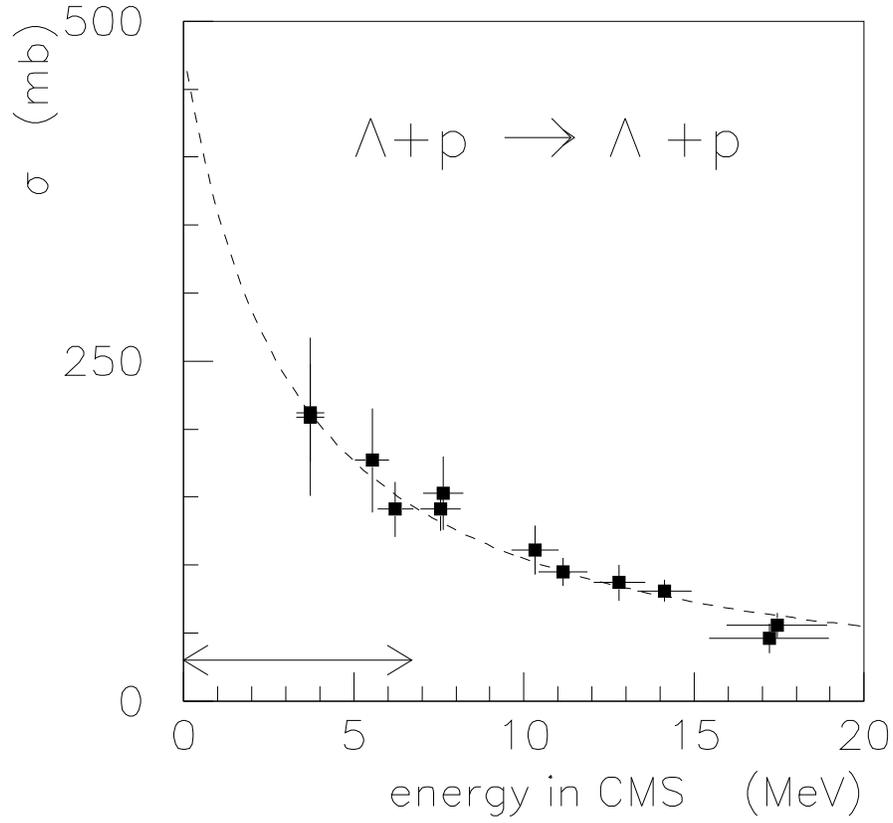}
\caption{Low energy $\Lambda p$ elastic scattering cross section as
measured in bubble chambers [1,2] as a function of the c.m.~energy 
$\varepsilon$. The arrow shows the range of energies covered by the present  
COSY-11 measurement.  The combined fit of our data with the scattering
results lead to $\gamma_1=-0.41$~fm$^{-1}$ and $\gamma_2=2.4$~fm$^{-1}$
and this is shown here as the dashed line.}
\label{elas_lp}
\end{figure}

In order to obtain values of $\bar{a}$ and $\bar{r}$ separately, we
must use extra experimental information such as for example $\Lambda$-$p$ 
elastic scattering cross section data. It is seen from Eq.~(1) that the 
normalisation of this at zero energy is proportional to the square of the 
scattering length. The low energy data of Ref.~\cite{Alex68,Sech68}
are shown in Fig.~\ref{elas_lp} together with a fit to the data on the basis of
Eq.~(1), where the parameters $\bar{a}$ and $\bar{r}$ are constrained
to lie on the maximum likelihood ridge of Fig.~\ref{cl_gg}. This is achieved
with $\bar{a}=-2.0$~fm and $\bar{r}=1.0$~fm, corresponding to 
$\gamma_1=-0.41$~fm$^{-1}$ and $\gamma_2=2.4$~fm$^{-1}$. The fit
therefore confirms that $\gamma_2\gg \gamma_1$ and that the $\Lambda p$ 
scattering data do not realistically allow for a separation between singlet
and triplet parameters. It is worth noting that our production data
are sensitive to much lower values of $\varepsilon$, as indicated by
the arrow, than the scattering data, and it is this region which
determines best the value of $\gamma_1$ and hence the position of the pole 
of the virtual bound state.

\section{Conclusions}

Through an analysis of the two-dimensional structure of the Dalitz
plot for the $pp\to pK^+\Lambda$
reaction at fixed energies within a simple final-state-interaction
model, we have established a strong constraint between the 
spin-averaged $\Lambda p$ scattering length and effective range. The data 
allow us to fit accurately the position of the spin-average virtual bound 
state $\gamma_1$. Since the data were taken at excess energies which
are inaccessible to low energy elastic scattering experiments, the
results are complementary and it is appropriate to make a combined fit
of the whole data set, leading to new values of $(\bar{a},\,\bar{r})$.

The total cross section data of Ref.~\cite{Bal97} have recently been analysed 
to determine values of $(\bar{a},\,\bar{r})$ \cite{Sibirtsev}. Their
argument is, however, somewhat cyclic since a final-state interaction
with fixed scattering length and effective range was already used in the
experimental analysis to fix the beam energy \cite{Bal97}. In
principle therefore these values should then be found by the fitting
procedure, though there are still of course the ambiguities discussed in
this paper. We avoid falling into this trap by not using the relative
normalisation of the event rate as a function of the beam energy
within our fitting procedure. It is rather the structure of the
two-dimensional Dalitz plots which fixes our parameter values.

Though we have implicitly assumed that the $\Lambda p$ system produced
from the near-threshold $pp \to p K^+\Lambda$ reaction is the same 3:1
spin-average seen in bubble chamber scattering experiments, this is
not guaranteed. Just as in the $K^-$-capture experiment \cite{Tan},
the basic reaction mechanism could favour the production of a
particular spin combination in the final state. If the present
experiment were extended through the use of a polarised beam and
target, then it would be possible to repeat the current analysis
separately in the singlet and triplet final states to separate these
important quantities.\\

\newpage
\noindent
{\large\bf Acknowledgements}

\noindent
We should like to thank  Dr~A.~Szczurek for very helpful discussions.
The research project was supported by the BMBF, the Polish Committee for
Scientific Research, the Bilateral Cooperation between Germany and
Poland, represented by the Internationales B\"uro DLR for the BMBF,
and the FFE program of the Forschungszentrum J\"ulich.
One of the authors (CW) is grateful to the FZ-J\"ulich for the 
consultancy which supported some of this work.\\[3ex]
%


\newpage

\newpage

\newpage


\newpage

\newpage

\newpage

\newpage

\end{document}